\pdfoutput=1
\documentclass[twocolumns,traditabstract]{aa}
\usepackage{graphicx}
\usepackage{amsmath}
\usepackage{amssymb}
\usepackage{latexsym}
\usepackage{graphicx}
\usepackage{gensymb}
\usepackage{xcolor}
\usepackage{enumerate}
\usepackage{txfonts}
\def\ltsima{$\; \buildrel < \over \sim \;$}
\def\simlt{\lower.5ex\hbox{\ltsima}}
\def\gtsima{$\; \buildrel > \over \sim \;$}
\def\simgt{\lower.5ex\hbox{\gtsima}}

\begin{document}
   \title{On the discovery of fast molecular gas in the UFO/BAL quasar APM 08279+5255 at z=3.912
      \thanks{
This work is based on observations carried out under project numbers S15CW and E15AF with the IRAM NOEMA Interferometer. IRAM is supported by INSU/CNRS (France), MPG (Germany) and IGN (Spain).
}
}

\author{C. Feruglio \inst{1,2}
\and
A. Ferrara \inst{2}
\and
M. Bischetti \inst{3,4,5}
\and
D. Downes \inst{6}
\and
R. Neri \inst{6}
\and
C. Ceccarelli \inst{7}
\and 
C. Cicone \inst{8}
\and
F. Fiore \inst{3}
\and
S. Gallerani \inst{2}
\and
R. Maiolino \inst{9}
\and
N. Menci \inst{3}
\and
E. Piconcelli \inst{3}
\and
G. Vietri \inst{3}
\and
C. Vignali \inst{10}
\and
L. Zappacosta \inst{3}
}

   \institute{ 
  INAF Osservatorio Astronomico di Trieste, Via G.B. Tiepolo 11, I-34143 Trieste, Italy   
\email{feruglio@oats.inaf.it}
   \and 
Scuola Normale Superiore, Piazza dei Cavalieri 7, I-56126 Pisa, Italy
\and 
INAF Osservatorio Astronomico di Roma, Via Frascati 33, 00078 Monteporzio Catone, Italy
\and 
Universit\'a degli Studi di Roma "Tor Vergata", Via Orazio Raimondo 18, I-00173 Roma, Italy
\and
Cavendish Laboratory, University of Cambridge, 19 J.J Thomson Ave., Cambridge CB3 0HE, UK 
\and 
   IRAM -- Institut de RadioAstronomie Millim\'etrique, 300 rue de la Piscine, 38406 Saint Martin d'H\'eres, France
\and
	Universit\'e Grenoble Alpes, IPAG, F-38000 Grenoble, France 
\and
 INAF Osservatorio Astronomico di Brera, via Brera 28, 20121, Milan, Italy
\and
	Kavli Insitute for Cosmology, University of Cambridge, Madingley Road, Cambridge CB3 0HA, UK
\and
Dipartimento di Fisica e Astronomia, Universit\'a degli Studi di Bologna, viale Berti Pichat 6/2, 40127, Bologna, Italy
             }


 
  \abstract{
 We have performed a high sensitivity observation of the UFO/BAL quasar APM 08279+5255 at z=3.912 with NOEMA at 3.2 mm, aimed at detecting fast moving molecular gas. We report the detection of blueshifted CO(4-3) with maximum velocity (v95\%) of $-1340$ km s$^{-1}$, with respect to the systemic peak emission, and a luminosity of $L' = 9.9\times 10^9 ~\mu^{-1}$ K km s$^{-1}$ pc$^{-2}$ (where $\mu$ is the lensing magnification factor). 
We discuss various scenarios for the nature of this emission, and conclude that this is the first detection of fast molecular gas at redshift $>3$.
We derive a mass flow rate of molecular gas in the range $\rm \dot M=3-7.4\times 10^3$ M$_\odot$/yr, and momentum boost $\dot P_{OF} / \dot P_{AGN} \sim 2-6$, therefore consistent with a momentum conserving flow. 
For the largest $\dot P_{OF}$ the scaling is also consistent with a energy conserving flow with an efficiency of $\sim$10-20\%. 
The present data can hardly discriminate between the two expansion modes. 
The mass loading factor of the molecular outflow  $\eta=\dot M_{OF}/SFR$ is $>>1$. 
We also detect a molecular emission line at a frequency of 94.83 GHz, corresponding to a rest frame frequency of 465.8 GHz, which we tentatively identified with the cation molecule $\rm N_2H^+$(5-4), which would be the first detection of this species at high redshift. We discuss the alternative possibility that this emission is due to a CO emission line from the, so far undetected, lens galaxy. Further observations of additional transitions of the same species with NOEMA can discriminate between the two scenarios.
}


\titlerunning{Fast molecular gas in APM 08279+5255}
\authorrunning{Feruglio et al.}
 \maketitle

\section{Introduction}

Molecular massive outflows driven by AGNs are today commonly detected in nearby galaxies, Seyferts and (U)LIRGs hosting AGN (Fiore et al. 2017 ad references therein), but not as such in the high redshift universe. 
Atomic ionised gas outflows do seem common at higher redshift, in z=2-3 quasars (e.g. Cano-Diaz et al. 2012, Carniani et al. 2015, Cresci et al. 2015, Harrison et al. 2014, Brusa et al. 2015, Bischetti et al. 2017).
Only a couple of high-redshift massive outflows of atomic gas are known, one seen in [CII] in the quasar J1148 at z=6.4 (Maiolino et al. 2012, Cicone et al. 2015), the other possibly seen in [NII] in the bright submillimetre galaxy HLSJ091828.6+514223 ay z=5.2 (Rawle et al. 2014).
Recent results suggest that [CII] outflows may also be present in starburst galaxies at $z\sim 5.5$ (Gallerani et al. 2017).  
What about molecular outflows? 
The most distant molecular outflows known to date are those found in the QSO RXJ0911.4+0551 at z=2.79 (Weiss et al. 2012), seen as a broad CO emission line, and in 
the Cosmic Eyelash at z=2.3 through detection of blueshifted OH absorption line (George et al. 2014). 
No observation of molecular outflows is available at earlier cosmic epochs. 

Two theoretical scenarios have been proposed to explain the expansion of an AGN-driven, nuclear wind into the interstellar medium of the host galaxy. 
One is the energy conserving wind scenario, whereby a 
nuclear semi-relativistic wind, with momentum $\dot P_{in} = v_{in} \times \dot M_{in} \approx L_{AGN }/c $, shocks against the ISM, and expands adiabatically pushing a cold wind on large scales ($\sim$kpc). 
This is the so called energy-conserving wind (King 2010, Zubovas \& King 2012, Faucher-Giguere \& Quataert 2012). 
The conservation of energy implies a momentum boost of the cold wind with respect to the nuclear wind of about $\dot P_{out}/\dot P_{in} \sim 10-20$.
The second is the momentum-conserving, or "gentle wind" scenario (Fabian et al. 1999). There, the nuclear semi-relativistic wind, pushed by the AGN radiation, cools rapidly while expanding in the ISM, and momentum and ram pressure of the nuclear wind are conserved during the expansion. This models predicts momentum boosts of a few ($\sim 5$).
To date the energy conserving scenario seems favoured based on few observations of molecular outflows in AGN host galaxies. 
Indeed, the only two cases where it was possible to measure both wind phases in the same galaxy, i.e. the nuclear semi-relativistic and the cold extended outflow (in Mrk 231 and IRAS F11119, Feruglio et al. 2015, Tombesi et al. 2015) measure momentum boosts of about 20. 
In addition, most molecular outflows in AGN host galaxies, which are mainly measured in nearby galaxies, show momentum boosts (measured in these cases by assuming $\dot P_{in} =L_{AGN }/c $), exceeding 10 (Fiore et al. 2017). Conversely, galaxy-wide ionised winds (traced mainly by [OIII]) are often momentum-conserving (Fiore et al. 2017).
More measurements are required to explore the momentum boost parameter space and assess the prevalence and relative importance of the two main wind physical models described above. 


To this aim, we undertook an observational campaign to explore whether a molecular wind is present in the broad absorption line (BAL) quasar APM 08279+5255, located at z=3.912, and to constrain its properties and energetics. 
We indeed consider this source as the next target, after Mrk 231 and IRAS F11119, to progress in our understanding of the lunching mechanisms of large scale winds, for several reasons.  
First, this BAL quasar is known for its strong and persistent semi-relativistic nuclear wind, observed with \emph{XMM-Newton} and \emph{Chandra} at several epochs (Saez \& Chartas 2011 and references therein). The highly ionised, ultra-fast nuclear outflow (UFO) is seen in absorption in the X-ray spectra, with a velocity of $\sim 0.3c$, and an outflow rate of a few tens solar masses per year. 
The quasar shows also a BAL system with velocity approximately $-2500$ km s$^{-1}$ as seen in [CIV] absorption at UV wavelengths (Irwin et al. 1998, Saturni et al. 2016).
Second, measurements of several CO rotational transitions from the quasar host galaxy are available (Downes et al. 1999, Lewis et al. 2002, , Weiss et al. 2007, Riechers et al. 2009, 2010), together with other atomic and molecular ones (Wagg et al. 2005, 2006, Garcia-Burillo et al. 2006).
Third, the emission is gravitationally magnified by factors ranging from a few to 100, depending on the adopted gravitational lens model, and observing frequency, which 
is an advantage for detecting broad, low surface brightness CO components with NOEMA. This, however, introduces also uncertainties due to the limited knowledge of the lensing model.  
Two main scenarios emerge from the high angular resolution studies performed from UV to cm wavelengths.
The first is the high-magnification scenario, which considers that the source is split into three images and has a large, but variable magnification factor as a function of the physical scales probed at the different observing frequencies, ranging from $\sim100$ in the nuclear regions (few to a few tens pc), to $\sim20$ at scales of $\sim$500 pc (Egami et al. 2000, Lewis et al. 2002, Krips et al. 2007, Oya et al. 2013). In addition, the variations detected in the NIR flux ratio of the source images imply that a microlensing event could have occurred  during the last decade (Oya et al. 2013). 
The second scenario proposes a 3-image lens model with moderate magnification factor (about 4 at all wavelengths) and a molecular disk seen close to face-on (Riechers et al. 2009). 
The lensing galaxy has never been identified, preventing us from discriminating between these two competing scenarios. 

In this paper we report about the most sensitive observation of molecular gas in APM 08279+5255 by using NOEMA. 
Section 2 describes observations. Section 3 discusses the detection of spectral features. In Section 4 we provide a discussion. Conclusions and future perspectives are presented in Section 5. 
Throughout the paper we adopt a $\Lambda$-dominated cosmology: $H_0=70$ km s$^{-1}$ Mpc$^{-1}$, $\Omega_\Lambda=0.7$, and $\Omega_m=0.3$.

\section{NOEMA observations}
The observations were designed to target the CO(4-3) emission line at a redshifted frequency of 93.856 GHz, and carried out under project number S15CW with the compact (D) 6 and 7 antenna array configurations of NOEMA.
Additional data were acquired under DDT project E15AF between April 5 and May 25, 2016, using both the B and the D array configuration with 7 antennas.

The data covers the spectral range from 454.85 to 472.53 GHz rest frame frequency. 
The quasar 0749+540 (0.4 Jy) was used as amplitude/phase calibrator. 
Absolute flux calibration is based on the MWC349 (1.12 Jy) and LKH$\alpha$ (0.24 Jy), whose fluxes are know with about 5\% accuracy at this frequency. 
The relative flux calibration between the different observations is better than 2\%. 

The on source time  is 33.4 hours (7-antenna equivalent time), after calibration, flagging of visibilities, and merging of all data. The achieved r.ms. noise is 96 $\mu$Jy/beam in 40 MHz channels (corresponding to 127.7 km s$^{-1}$), and 10 $\mu$Jy/beam in the continuum on the full 3600 MHz bandwidth, which makes it the most sensitive observation of APM08279+5255 at mm wavelengths.
Natural weighting and a simple cleaning algorithm (\textit{hogbom}) were used to image the combined data set, in order to minimise secondary lobes.
The resulting synthesised beam is $4.0\times 3.6$ arcsec, PA 49 deg. 

\section{Results}

\subsection{Continuum emission}
\label{c}
The observed frame 3.2 mm continuum (i.e. 650 $\mu$m rest frame) was estimated by averaging the visibilities in spectral windows. We tested different spectral windows in ranges free of lines with consistent results.
A point source fit to the visibilities averaged over all spectral windows gives a continuum flux density of $1.1 \pm 0.02$ mJy. 
Considering an additional systematic uncertainty of $\sim5\%$,  
it is in agreement with previous measurements (Downes et al. 1999, Weiss et al. 2007). 
The rest frame 650 $\mu$m continuum emission is spatially unresolved (Table \ref{tablines}, Figure 1 where the synthesized beam is shown in the inset). In particular we do not resolve the two images (North-East, NE, and South-West, SW) of the continuum found by Krips et al. (2007) using the SMA (which are separated by about 0.3$"$). 

\begin{figure}
\centering
\includegraphics[width=7cm]{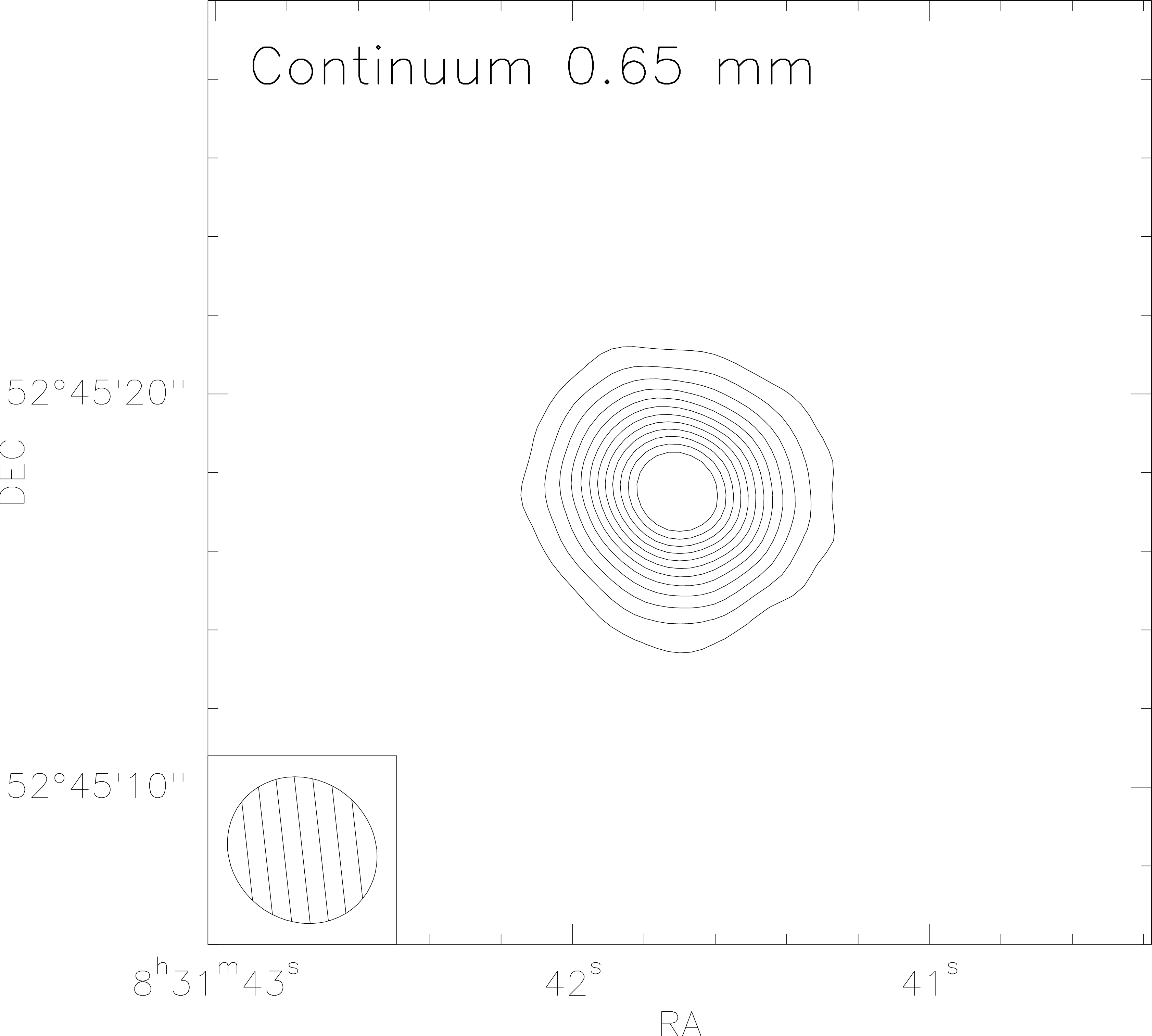}
\caption{Map of the 650 $\mu$m rest frame continuum emission (3.2 mm observed frame), averaged over 38 emission-line free, 40-MHz wide channels (i.e. a bandwidth of 1520 MHz). First level step is 5$\sigma$, increasing by 5$\sigma$, $\sigma= 15$ $\mu$Jy/beam.
The synthesised beam ($4.0\times 3.6$ arcsec, PA 49 deg) is shown in the inset.} 
\label{continuum}
\end{figure}

\begin{figure*}[t]
\centering
\includegraphics[width=\textwidth]{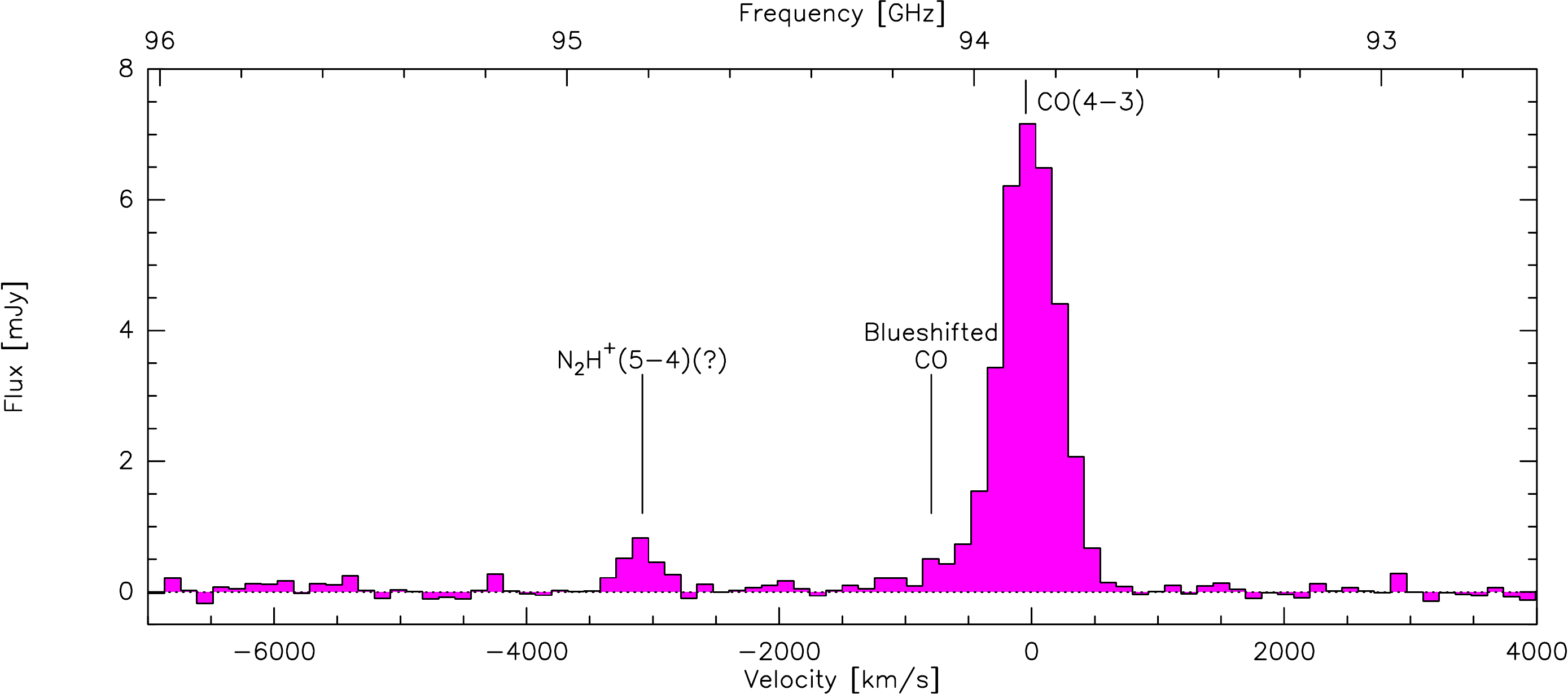}\\
\caption{The full-bandwith spectrum of APM 08279 in the region around CO(4-3). The spectrum was spatially integrated over the source, after continuum subtraction, 
over a polygon around the source, determined by the 2$\sigma$ contour level in a line map integrated over the velocity range 
$-1340$ to +670 km s$^{-1}$.
The observed frequency is reported on the upper x-axis. 
The continuum has been subtracted in the uv plane as detailed in Sec. \ref{coline}. Labels marking spectral features, are just to guide the eye.
Actual line identifications are discussed in Section 4.
}
\label{spectrum}
\end{figure*}

\subsection{CO(4-3) emission line}\label{coline}

Figure 2 shows the spectrum of the CO(4-3) spectral region.
To obtain this spectrum, we first subtracted the continuum, and then made  
a broad-band line map integrated over the velocity range $-1340$ to +670 km s$^{-1}$, with respect to the CO emission peak at 93.856 GHz, corresponding to a redshift of z$=$3.912, consistent with previous PdBI CO and HCN observations (Weiss et al. 2007).
On this broad-band line map, we traced a polygon at the location of 
the 2$\sigma$ contour around the velocity-integrated line source.
We then returned to the individual channel maps, and spatially integrated each spectral
channel over this same polygon. The spectrum shows the result of the spatial integration for
each spectral channel.
On the blue side of CO(4-3) we detect emission out to velocities exceeding $-1000$ km s$^{-1}$  with respect to the CO peak (Fig. 2 and 3). 
The spectrum shows some positive residuals in the continuum in the bluest part of the spectrum, at frequency $\sim96$ GHz. 
These may be due to a slope of the continuum, which is not accounted for when using the averaged-visibility method to subtract it. 
This putative slope can hardly affect the estimate of the main line CO(4-3), but may affect the faint wing of the line. 
In order to account for this slope, 
we extract the spectrum from the clean cube using a mask that encompasses the emission with a threshold of 2$\sigma$ (this corresponds to a area of about $5\times6$ arcsec), and fit it with a combination of a linear function, and multiple Gaussian functions for the continuum and the emission lines. The best fit parameters are reported in Table \ref{tabgauss} . 
We find a normalisation of the continuum of $1.10\pm0.02$ mJy, and a slope of  $-6.1 \pm 5.2 \times10^{-6}$ mJy s km$^{-1}$, by fitting the entire bandpass (the fit is very good without any strong residuals). 
We find that two Gaussian components are required to best fit the CO(4-3), one at $-12\pm5$  km s$^{-1}$, with peak intensity S=$7.7\pm0.1$ mJy and $FWHM=540\pm12$ km s$^{-1}$, plus a blueshifted component at $-800$ km s$^{-1}$ and with $FWHM=640^{+750}_{-330}$ km s$^{-1}$. This fit is performed in the velocity range from $-5000$ to 4000 km s$^{-1}$, with 70 points, 58 degrees of freedom, and has a $\chi^2=102.9$.  
For comparison, a fit of the CO(4-3) line with a single Gaussian function gives a $\chi^2=149.5$ (61 degrees of freedom).
Errors are given at the 90\% level for one parameter of interest. 
Figure 3 shows the spectrum, together with the best fit continuum and line components. 
We find integrated intensities of $I(CO)_{sys} = 4.4\pm0.1$ Jy km s$^{-1}$ for the systemic CO, and  $I(CO)_{blue}=0.25^{+0.25}_{-0.07}$ Jy km s$^{-1}$ for the blueshifted component, and thus a ratio $I_{sys}/I_{blue}=18$. 
There is no evidence for a redshifted fast component.

\begin{table*}
\caption{Measured parameters for lines and continuum derived from fits of the visibilities.}
\begin{center}
\begin{tabular}{lccc}
\hline
\multicolumn{1}{l} {Parameter}&
\multicolumn{1}{c} {CO(4-3) systemic}&
\multicolumn{1}{c} {CO(4-3) blue}&
\multicolumn{1}{c} {$\rm N_2H^+(5-4)$} \\
& $[-700,670]$ km s$^{-1}$  & $[-1340,-700]$ km s$^{-1}$ & $[-3540,-2662]$ km s$^{-1}$ \\
\hline
Emitted frequency (GHz) &   461.0408 &  -   & 465.732   \\ 
Observed frequency (GHz) &  93.856   &  -   & 94.85    \\
RA  [08:31:] &    $41.70 \pm 0.01$ &   $41.7 \pm 0.1$  &  $41.7 \pm 0.1$ \\
DEC [52:45:] &    $17.52 \pm 0.01$ &   $16.40 \pm 0.1$  &  $17.5 \pm 0.1$  \\
$S$ (mJy)    &    $3.02\pm0.04$    &   $0.43 \pm 0.08$ &  $0.337\pm0.035$   \\
F.W.H.P. (arcsec)   & $0.65\pm0.04$      & - & - \\
I (Jy km s$^{-1}$) & $4.1\pm0.05$   &  $0.27\pm0.05$   & $0.30\pm0.03$ \\
\hline
\end{tabular}
\end{center}
Notes. The table gives statistical errors. Additional systematic uncertainties are about 5\%. Additional 1$\sigma$ uncertainty on the positions can be derived from the Reid et al. (1988) relation,  $\Delta \theta (arcsec) = \sigma_{beam} / (2\times SNR)$.
Parameters are derived from point or circular Gaussian function fit of the visibilities averaged in the corresponding velocity ranges.
\label{tablines}
\end{table*}

\begin{table*}
\centering
\caption{Measured parameters for lines and continuum derived from Gaussian fit of the spectrum.}
\begin{tabular}{lccccc}
\hline
\hline
Component &  Peak   & S                     & FWHM     & I                                & L' $\mu^{-1}$  \\
                    &  [km s$^{-1}$] & [mJy]               & [km s$^{-1}$]       & [Jy km s$^{-1}$]                  & [K km s$^{-1}$ pc$^{-1}$]  \\
\hline 
Systemic CO &   $-12\pm5$ &  $7.7\pm0.1$  & $540\pm12$    & $4.4\pm0.1$  &    $1.75 \times 10^{11}$     \\
Blueshifted CO &   $-800^{+360}_{-100}$ & $0.37^{+0.15}_{-0.1}$  &  $640^{+750}_{-330}$ & $0.25^{+0.25}_{-0.07}$   & $9.9\times10^{9}$ \\
94.83 GHz & $-3100\pm25$  & $0.84^{+0.14}_{-0.12}$ &  $340 \pm 50$ & $0.31\pm0.05$ &    $1.2\times 10^{10}$   \\ 
\hline
 & & & & &  \\
 \hline
                   &         & S [mJy]  & Slope & & \\  
Continuum &    -    & $1.10\pm0.02$ & $-6.1\pm 5.2 \times 10^{-6}$    &  - &  - \\
\hline
\end{tabular}\\
Notes. Errors are given in 90\% level for one parameter of interest. $\mu$ is the lensing magnification factor.
\label{tabgauss}
\end{table*}

\begin{figure}[t]
\centering
\includegraphics[width=\columnwidth]{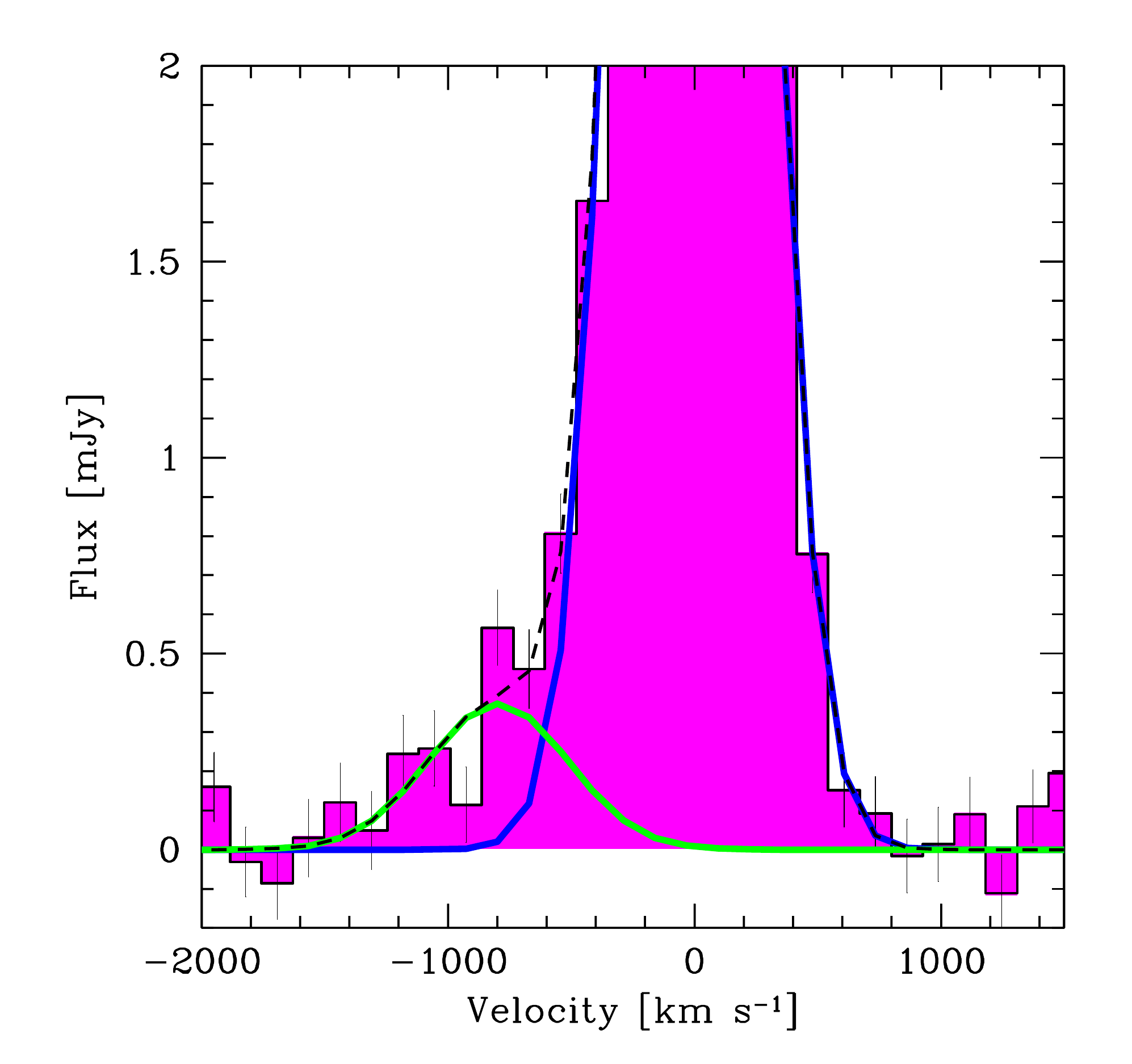}
\caption{Zoom-in view of the spectrum around CO(4-3), integrated in a polygon set by the $>2\sigma$ noise level around the QSO. A double gaussian fit has been applied to CO(4-3), including a systemic (blue) and a blueshifted (green) line component. The dashed line shows the combination of the two gaussian functions.
The continuum has been fitted with a linear function and subtracted as detailed in section 3.2.}
\label{fit}
\end{figure}

\begin{figure*}[t]
\centering
\includegraphics[width=16cm,angle=0]{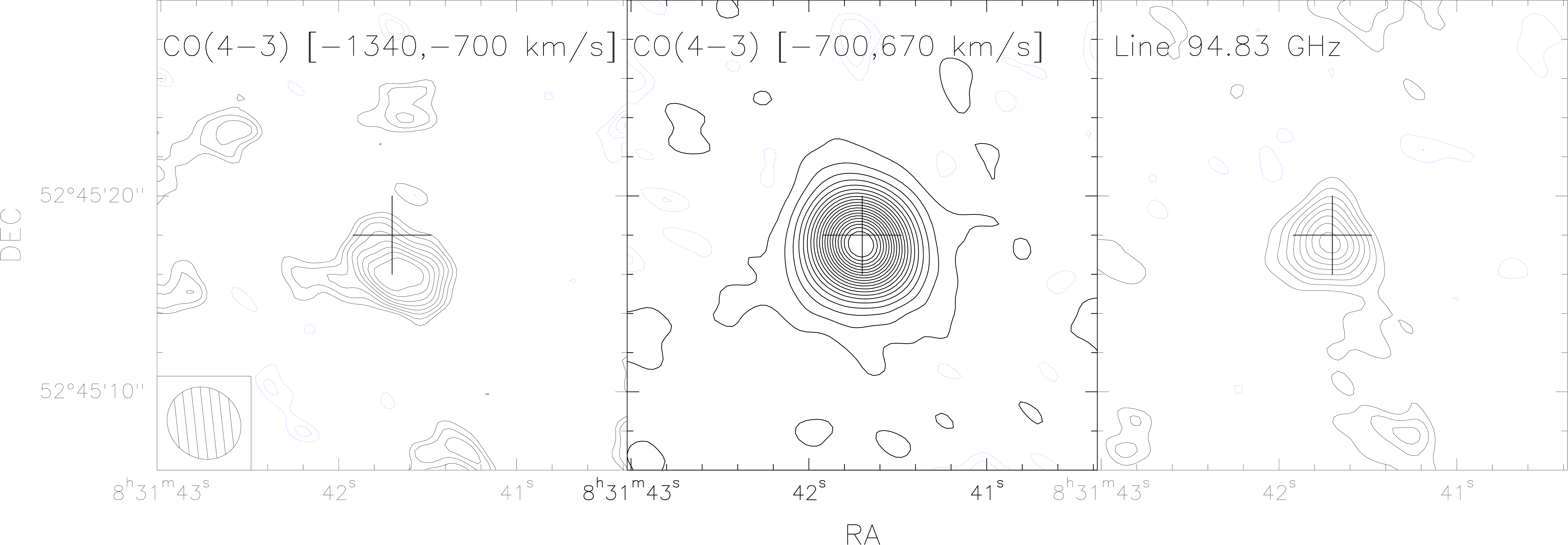}
\caption{
Left panel: the blueshifted CO component is integrated between $peak_{blue}-2\sigma_{blue}=-1340$ km s$^{-1}$ and $CO_{sys}-3\sigma_{sys}=-700$ km s$^{-1}$  (i.e. according to the  Gaussian fit in Table \ref{tabgauss}). 
Levels start at $\pm2 \sigma$, increase (decrease) by 0.5$\sigma$ ($\sigma=39$ $\mu$Jy/beam for an integrated bandwidth of 240 MHz$=$767 km s$^{-1}$).
Middle panel: the CO systemic line integrated in the range $Line-peak\pm3\sigma$. Levels start at $\pm 2 \sigma$, increase (decrease) by 5$\sigma$ ($\sigma=27.7$ $\mu$Jy/beam for 480 MHz (=1533 km s$^{-1}$) integrated bandwidth).
Right panel: the line at 93.85 GHz integrated in the range $Line-peak\pm3\sigma$.
Levels start at $\pm2 \sigma$, increase (decrease) by 1$\sigma$ ($\sigma=34$ $\mu$Jy/beam per 320 MHz =1022 km s$^{-1}$ integrated bandwidth). 
The cross denotes the phase tracking centre and is $4\times4$ arcsec wide.
}
\label{co}
\end{figure*}

Maps of the spectral lines components are shown in Figure \ref{co}. The spectral ranges of integration for each component  have been chosen according to the best fit peak position and 
$\sigma$ of the multi-gaussian fit of the spectrum. 
The blueshifted CO emission is detected at 5.5$\sigma$. 
The emission line at 94.83 GHz is discussed in Section \ref{uif}.
We have also performed visibility fitting of the spectral components, using the integrated maps in Fig. \ref{co} (Table \ref{tablines}). 
For the emission lines we have averaged the visibilities in velocity ranges according to the peak and $\sigma$ best fit parameters of the spectrum (see Table \ref{tablines}). 
We have used point source models for all components but the systemic line, which is best fitted by a
circular gaussian function. 
The integrated intensities from visibility fit agree with those derived from the fit of the spectrum. 

We derive in the following the line luminosities and molecular gas masses, using the results reported in Table \ref{tablines}. 
The systemic CO(4-3) component has a apparent luminosity $L'(CO) = 1.75 \times 10^{11}~\mu^{-1}~ \rm K~km~ s^{-1}~pc^{-2}$, where $\mu$ is the lensing magnification factor, in agreement with Downes et al. (1999), and with Riechers et al. (2009).
For the blueshifted component we measure $L'(CO)_{blue}=9.9 \times 10^{9}~\mu^{-1}~ \rm L_{\odot}$, where $\mu$ is the lensing magnification factor of the fast moving gas. 
We note that in principle the magnification factors of this component and that of the main CO component may be different, as the molecular disk and high velocity component may have different physical sizes. 

\begin{figure*}[t]
\centering
\includegraphics[width=7cm]{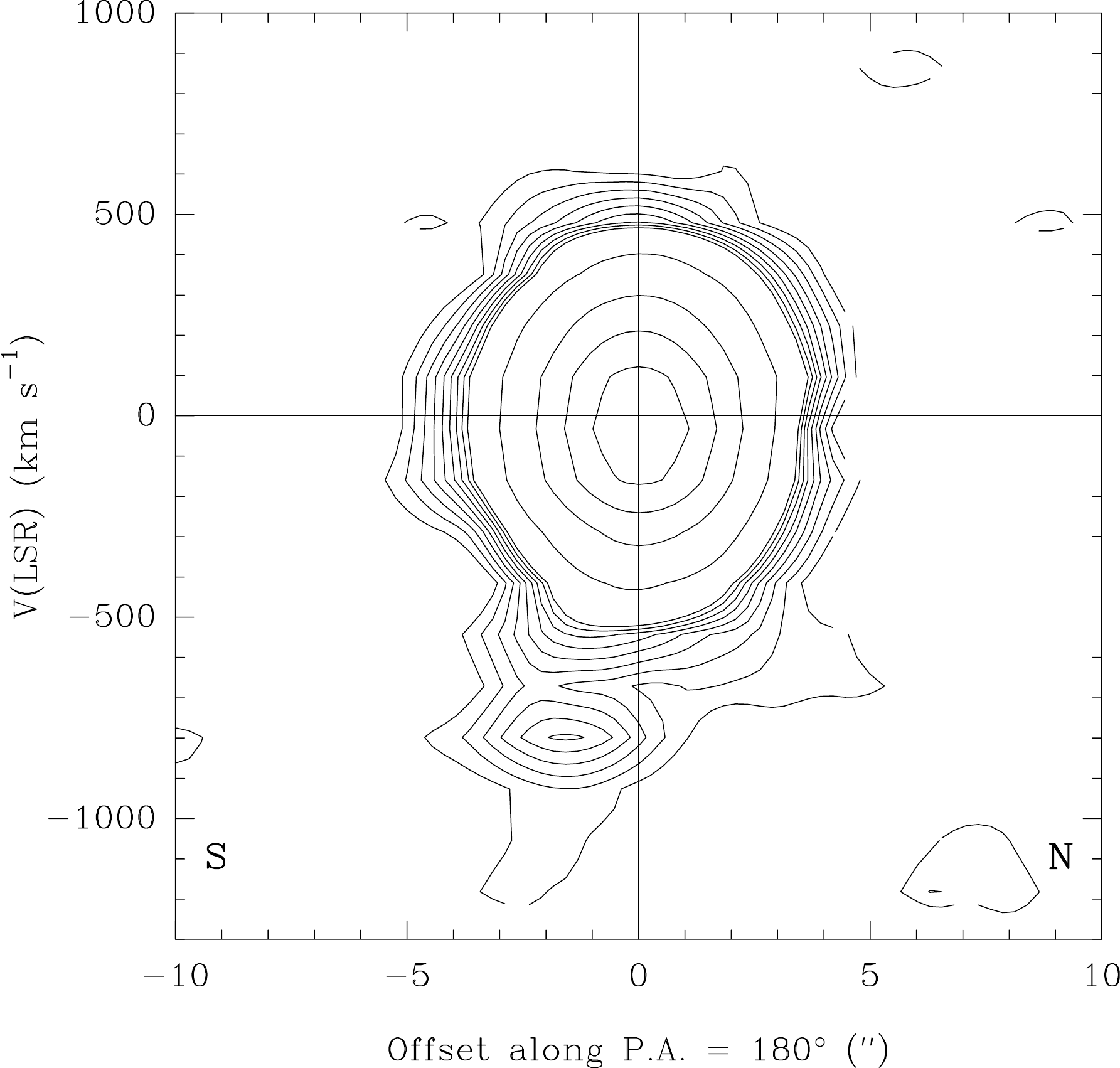}
\includegraphics[width=7cm]{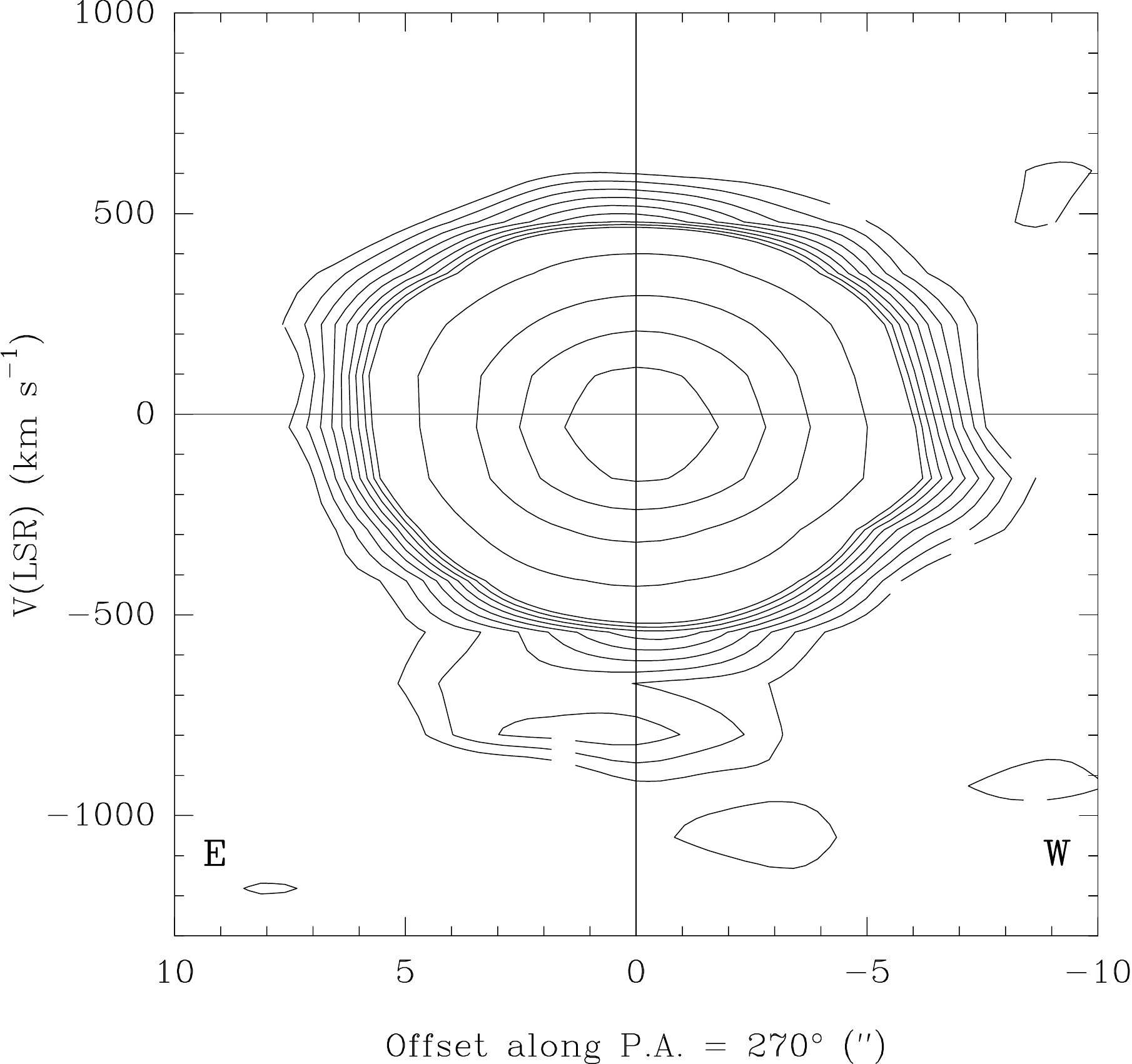}
\caption{
Position-velocity diagrams of the CO(4-3) emission line. Left panel: PV plot through the line connecting CO(4-3) peak position to the high velocity CO peak position, corresponding to a PA =180 deg (i.e.  South-North direction). Right panel: PV plot along a PA = 270 deg (West-East). 
Levels are from  2 to 10$\sigma$ by 1$\sigma$, 20, 40, 60, 80 and 100$\sigma$, $\sigma = 96~\mu$Jy/beam for 40 MHz. Each slice is 1 arcsec thick.}
\label{pv}
\end{figure*}

Figure \ref{pv} shows position-velocity (PV) diagrams of the CO(4-3) emission line, along the direction of the maximum separation between the narrow- and the wing-emission (which corresponds to the south-north direction, or PA of 180 deg, PA positive from North counterclockwise ); and in a direction orthogonal to this. 
Both PV diagrams have been reprojected to be centred on the QSO position 08:31:41.7, 52:45:17.50. 
The blue wing emission peaking at $-800$ km s$^{-1}$ is located at $1.5\pm0.15"$ south of the QSO. 
However, differential magnification can occur and an observed projected separation of 1.5"" may translate to a smaller physical separation due to the lensing. Therefore this emission can occur anywhere between $\sim$10 kpc (projected distance) and the central parts of the QSO host galaxy.
No emission is detected in the channels above 500 km s$^{-1}$ on the red side of the line.

\subsection{Emission line at 94.83 GHz}\label{uif}

We detect an emission line with peak at 94.83 GHz, corresponding to rest frame 465.8 with z=3.912 (Figure 1 and Figure 4, right panel).
The fit of the spectrum with a Gaussian function gives a peak intensity of $0.84^{+0.14}_{-0.12}$ mJy, a $\rm FWHM=340\pm50$ km s$^{-1}$, centred at $-3100$ km s$^{-1}$ with respect to the CO peak (Table \ref{tabgauss}). 
The integrated intensity is $0.31\pm0.05$ Jy km s$^{-1}$.
By averaging the visibilities in the spectral range 94.76 to 94.94 GHz, and by fitting them with a point source model, 
we derive an integrated intensity of $0.30\pm0.03$ Jy km s$^{-1}$, consistent with the Gaussian fit of the spectrum (Table \ref{tablines}). 
The identification of this emission line is discussed in Section 4.2.

\begin{table*}
\caption{Molecular outflow parameters.}
\begin{center}
\begin{tabular}{lccccccc}
\hline
\multicolumn{1}{l} {Model}&
\multicolumn{1}{c} {$\mu$}&
\multicolumn{1}{c} {R}&
\multicolumn{1}{c} {M(H$_2$)}&
\multicolumn{1}{c} {$\rm v_{max}$}&
\multicolumn{1}{c} {$\rm \dot M_{OF}$}&
\multicolumn{1}{c} {$\rm \dot P_{OF}$} &
\multicolumn{1}{c} {Refs.$^a$} \\
&     &  [kpc] & [$M_\odot]$  & [km s$^{-1}$] &   [$\rm M_\odot /yr]$ & $\rm [dyn]$ &     \\
\hline
model1 &  20 & 0.270 & $1.98\times10^8$  & $-1340$ & $3.0\times10^3$   & $2.5\times10^{37}$   & (1)  \\
model2 &  4   & 0.550 & $9.9\times10^8$    & $-1340$ & $7.4\times10^3$   & $6.3\times10^{37}$   & (2) \\
\hline
\end{tabular}
\end{center}
\label{taboutflow}
Notes. $^a$ References for the magnification factor and the size of the molecular outflow, here assumed equal to the size of the molecular disk, are: (1) Downes et al. 1999, (2) Riechers et al. 2009.
\end{table*}

\begin{table}
\caption{X-ray nuclear wind (UFO) physical parameters.}
\begin{center}
\begin{tabular}{lccccc}
\hline
\multicolumn{1}{l} {Model}&
\multicolumn{1}{c} {$\mu$}&
\multicolumn{1}{c} {$v_{UFO}$}&
\multicolumn{1}{c} {$\dot M_{UFO}$}&
\multicolumn{1}{c} {$\dot P_{UFO}$} &
\multicolumn{1}{c} {Refs.} \\
&      & [km s$^{-1}$] &   [$\rm M_\odot /yr]$ & $\rm [dyn]$ &     \\
\hline
model1 &  100 & 0.16-0.36c & $8-12$   & $1.0\times10^{37}$  &  1 \\
model2 &   1    & 0.22c & $7.7-38^{(a)}$   & $3.2-16\times10^{36}$  &   2\\
\hline
\end{tabular}
\end{center}
\label{tabufo}
Notes. $^{(a)}$ for a black hole mass of $2\times 10^9-10^{10}$ M$_\odot$.  
References: (1) Saez \& Chartas (2011), (2)  Hagino et al. (2017).
\end{table}

\begin{figure}[t]
\centering
\includegraphics[width=\columnwidth]{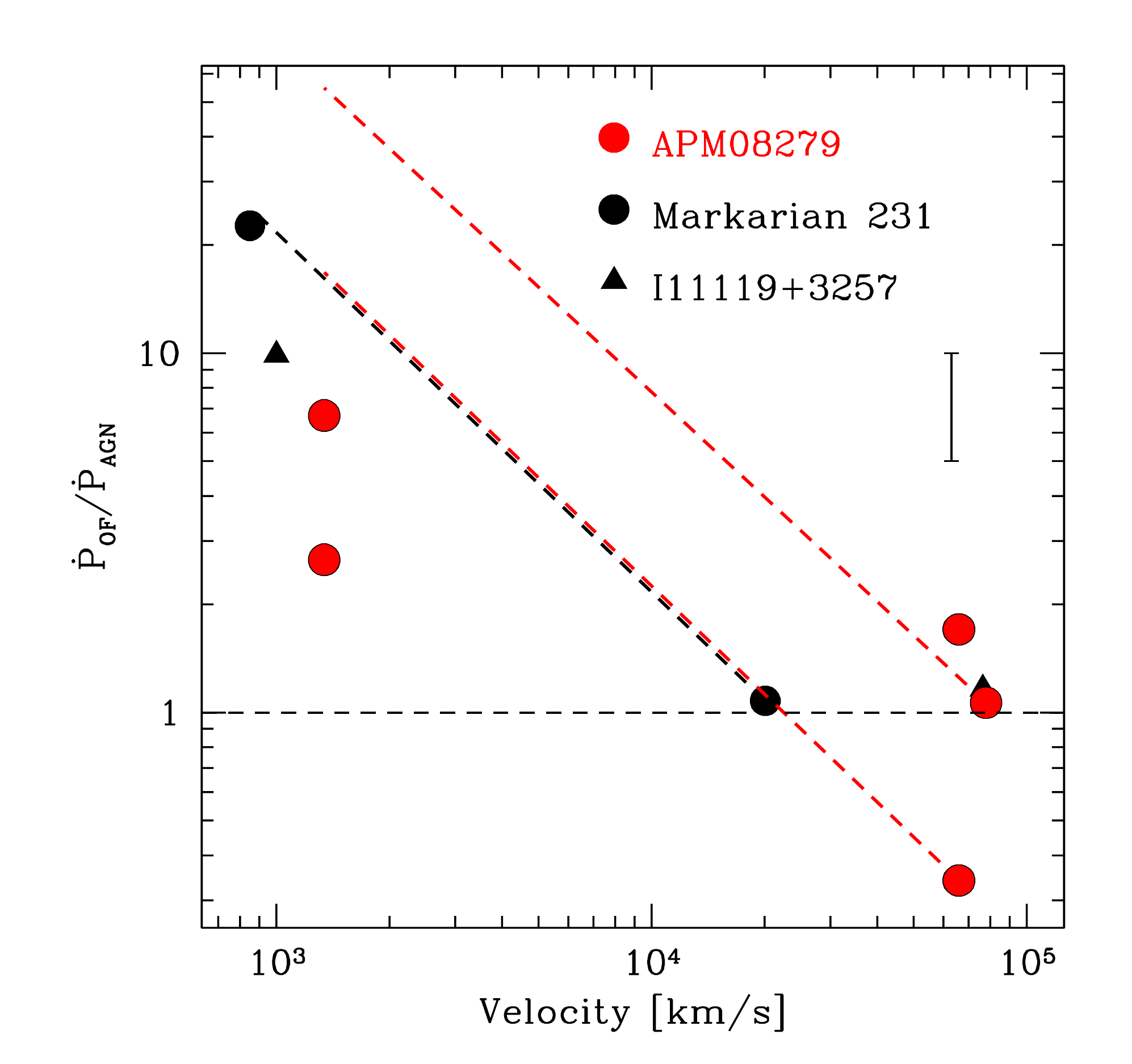}\\
\caption{$\dot P_{OF}/\dot P_{AGN}$ of the molecular and of the nuclear winds (UFOs) versus the wind velocity for APM08279 (red symbols), Mrk231 (black circles, from Feruglio et al. 2015), and IRAS11119 (black triangle, from Tombesi et al. 2015). 
UFOs are at $>10^4$ km s$^{-1}$. Molecular outflows are at velocities $\sim 1000$ km s$^{-1}$, for each source. 
For APM08279 the two values of $\dot P_{OF}$ listed in Table 3 are used to derive $\dot P_{OF}/\dot P_{AGN}$. The oblique dashed lines show the expectation for energy conserving flows, the horizontal dashed line shows that for momentum-conserving flows. The errorbar shows the typical statistical error.}
\label{boost}
\end{figure}

\section{Discussion}

\subsection{Nature of the CO(4-3) blueshifted emission}

We can identify four scenarios for the origin of the blueshifted CO(4-3) emission:
(1) emission from molecules with rest frame transitions close to that of the CO(4-3) line;
(2) a separate merging galaxy; 
(3) inflowing molecular gas from behind the QSO, and 
(4) molecular gas outflowing from the front side of the QSO.

To explore the first possibility, we refer to the spectral scan of the  Orion-KL hot cloud core in the 455-507 GHz frequency range, provided by White et al. (2003). 
 A better region to compare line ratios with would be Sgr B2, which is likely to have excitation and chemistry more similar to APM08279 than Orion KL. Unfortunately, there is no information on this spectral region of Sgr B2 published so far. We caution however, that the conditions in APM08279 may be very different from any Galactic molecular region.

The blueshifted emission is centred at 94.108 GHz (i.e. around $-800$ km s$^{-1}$ from the CO peak, Table 1), corresponding to rest frame 462.258 GHz, for z=3.912.
According to our Gaussian fit the peak line intensity ratio of CO(4-3) systemic/blueshifted is about 21. 
According to White et al. (2003), around that frequency we expect to find 
$\rm^{34}SO$ at 462.236 GHz, and $\rm^{13}CS(10-9)$ at 462.334 GHz.
In Orion KL CO(4-3)/$\rm^{34}SO$=4, and $\rm CO(4-3)/ ^{13}CS(10-9)=12$ (line peak). 

We extract a spectrum around the position of the blueshifted CO emission (as given in Table 1), within a region encompassed by the 2$\sigma$ noise level. 
A Gaussian fit to this spectrum yields a ratio CO(4-3) peak / blue-wing $\sim 15$. 
Based on this reasoning, $\rm^{34}SO$, $\rm^{13}CS(10-9)$ or a blend thereof, could contribute to the blue wing.

 About the merging scenario, we note that the merging galaxy may or may not be gravitationally lensed, depending on the configuration of the system. In the following we assume that it is gravitationally lensed by the same $\mu$ factor for both the dust and the gas components. 
The putative merging galaxy has not been detected at any wavelength (Ibata et al. 1999, Krips et al. 2007, Oya et al. 2013). 
In particular, the sub-mm continuum of this component is undetected in SMA data (down to 1$\sigma$ sensitivity of 1.7 mJy/beam at 200 $\mu$m rest frame, Krips et al. 2007). 
We can rely on a typical gas/dust mass ratio to verify whether the SMA upper limit in the continuum is useful. The molecular gas mass is $9\times 10^9 \mu^{-1}\rm~ M_\odot$, by assuming a MW conversion factor $\alpha_{CO}=4.6~ \rm K~ km~ s^{-1} ~pc^2)^{-1}~ M_{\odot}$, and a ratio CO(4-3)/CO(1-0) = 5 (Daddi et al. 2015, Table 2). 
The gas to dust mass ratio in APM08279 is about 200 (M$_{dust} =5\times 10^8 \mu{^-1}~\rm M_\odot$ , Downes et al. 1999, Weiss et al. 2007) - however in APM the QSO has been shown to strongly overheat the dust compared to a galaxy so this ratio may not be appropriate for any galaxy. 
We can use a reference value of M$_{gas}$/M$_{dust} = 100$ (BzK-21000 galaxy at z=1.5, Magdis et al. 2011; Scoville et al. 2012).
If we applied this ratio to the candidate merging galaxy, we would expect a dust mass of the order 
M$_{dust}= 9\times 10^7 \mu^{-1}~\rm M_\odot$.  
 If we assume the galaxy has the same properties as a BzK-21000, by scaling by the redshift, and assuming equal gravitational boost factor for the gas and for the dust, this translates into a 200 $\mu$m flux of 0.2 mJy. This is well below the detection limit of the SMA data of Krips et al. (2007). 
 
 On the other hand, our NOEMA data have exquisite sensitivity but they do not have the angular resolution to locate the continuum emission. Location is possible in the velocity space, see PV diagrams, but the continuum emission would be contained in the NOEMA beam of 4x3.6\arcsec. 
Detection of higher J transitions (e.g. CO(6-5) and CO(9-8)), together with detection and location of the continuum, are needed to constrain the excitation of the gas and the nature of the emission.

The third possibility is inflowing molecular gas from behind the QSO. This cannot certainly be a nuclear inflow of gas accreting onto the galactic nucleus, because of all known nuclear inflows occur with low velocities of a few tens to 100 km s$^{-1}$ at most (Querejeta et al. 2016; Combes et al. 2013). One other possibility is that of a large scale inflow, that is accretion of molecular gas onto the galaxy on larger physical scales (e.g. along one or more cosmic filaments, or from the circum-galactic medium, CGM), which can in principle occur with larger velocity. This would be the first time an inflow of molecular gas from the CGM is observed.

The last possibility is outflowing molecular gas from the front side of the QSO, and it is discussed in the next section.

\subsection{Molecular outflow}

We discuss in the following the properties of a molecular outflow, as traced by the blueshifted CO(4-3) emission. 

We recall that two main scenarios were proposed for the lensing model of APM. 
One requires a high magnification factor, variable depending on the physical scale (about $\mu=100$ for the nuclear region, sub-pc out to a few tens pc), and decreasing for larger spatial scales (Egami et al. 2000, Krips et al. 2007, Chartas et al. 2009, Downes et al. 1999). 
The other requires magnification factor of a few, and constant at all scales (e.g. Riechers et al. 2009). 
This uncertainty on the magnification factor also reflects in a factor of $\sim 10$ uncertainty on the bolometric luminosity, which ranges from a few $10^{47}$ to a few $10^{48}$ erg/s (Irwin et al. 1998, Riechers et al. 2009, Saturni et al. in prep.). 
We will discus the results in the framework of the two models described above.  
 
From the systemic CO component we derive a molecular gas mass of $\rm M(H_2)= 1.1  \times 10^{11}~\mu^{-1}~\rm M_{\odot}$, by adopting a ratio CO (4-3)/(1-0) = 20 (Weiss et al. 2007), and $\alpha_{CO}=0.8~ (\rm K~ km~ s^{-1} ~pc^2)^{-1}~ M_{\odot}$ (Downes \& Solomon 1998), to convert luminosity into mass.  

In order to convert the luminosity of the blueshifted CO component into a mass of molecular gas, we adopt a conversion factor of $\alpha_{CO}=0.5~ \rm K~ km~ s^{-1} ~pc^2)^{-1}~ M_{\odot}$. 
This has often been used to derive masses of outflowing gas (e.g. Feruglio et al. 2010, Cicone et al. 2014), based on the fact that this value has been measured in the molecular outflow in M82 (Weiss et al. 2001). 
This $\alpha_{CO}$ has often been considered as a safe lower limit on the mass of high velocity gas, because it is generally considered unlikely, although possible, to have smaller conversion factors (see, e.g. Dasyra et al. 2016).
Under this assumption, the lower limit on the molecular gas mass in the outflow would be $\rm M(H_2)_{of}=3.96 \times 10^{9}~\mu^{-1}_{of}$  $\rm M_{\odot}$. 

The molecular gas outflow rate is computed as follows:

\begin{equation}
\dot M_{of} = 3 \times \frac {v_{max,of} \times M(H_2)_{of} }{R_{of}} 
\end{equation}

where $v_{max,of}$ is the maximum outflow velocity, $\rm M(H_2)$ is its mass, and $R_{of}$ is the radius of the region reached by the outflow. 
The data do not allow to measure the sizes of the emitting regions, therefore we must make some assumptions. 
Two scenarios are discussed in the following, their main parameters being resumed in Table 3. 
In the first scenario (model1) we assume that the fast moving gas has the same size as the molecular disk, as measured by Downes et al. (1999). They 
found a radius of the molecular disk of 270 pc for a $\mu\sim20$. We derive in this case a outflow rate of $3.0\times 10^3~M_\odot/yr$. 
The second model  (model2) adopts a disk size of 550 pc and a $\mu=4$ (Riechers et al. 2009), from which we estimate a outflow rate of $7.4\times 10^3~M_\odot/yr$ (Table 3). 
The corresponding momentum fluxes,  $\dot P = v_{max} \times \dot M$, are also reported in Table 3.

Interestingly, we do not find evidence for fast, redshifted gas. Although, as noted in section 4.1, it is possible that the blueshifted CO traces inflowing material from the back side of the QSO, it can also suggest an asymmetric outflow with only a blueshifted blob of outflowing material from the front side. We note that the blue and redshifted components in molecular outflows do not have to be symmetric. For example in Mrk 231 there is a factor of $\sim 2$ in brightness between the red and blue wings of all observed CO transitions (J=1-0, 2-1 and 3-2), therefore the non detection of the redshifted component in APM 08279 may be a sensitivity effect. 

We discuss in the following the wind momentum boost. 
About the nuclear UFOs, two main scenarios have been proposed. That proposed by Saez \& Chartas (2011), displays  a large magnification factor (about 100), two outflow components with velocities 0.16c and 0.36c, and a total outflow rate of $\dot M_{ufo} = 21~ M_\odot/yr$. 
The second scenario, recently proposed by Hagino et al. (2017) is based on a physical model where the outflow rate scales as 
\begin{displaymath}
\dot M_{UFO} = 10.5 \frac{M(BH)}{2\times10^9}  \frac{v_{UFO}}{0.3c}~ M_{\odot}/yr
\end{displaymath}

For $v_{UFO}=0.22c$, and a black hole mass in the range $M(BH)=2 \times 10^9 - 10^{10}$ M$_\odot$ (Hagino et al. 2017 and references therein), we estimated a nuclear outflow rate of 7.7-38 $M_{\odot}/yr$. The main parameters of the UFO are reported in Table 4. 

The momentum boost, defined as $\dot P_{OF} / \dot P_{AGN} = \dot P_{OF} / (L_{AGN}/c)$, is plotted in Figure 6.
We adopt a bolometric luminosity of $10^{47.5}$ erg/s (we recall that the uncertainty of $L_{bol}$ may be as big as a factor of $\sim$10).
Note that the three estimates of $\dot P_{UFO}/ \dot P_{AGN}$ bridge the unity value, expected for a momentum conserving nuclear wind.

For the molecular outflow we derive $\dot P_{OF}/ \dot P_{AGN}=2-6$, i.e.  within the range of values compatibles with a momentum conserving flow, in geometrically thin and extremely optically thick expanding shells (Thompson et al. 2015). 
For the largest $\dot P_{OF}$ the scaling implies a energy conserving flow with an efficiency of $\sim$10-20\%. 
Otherwise, for the largest values of $\dot P_{UFO}$ and the lowest of $\dot P_{OF}$ the scaling is consistent with a momentum conserving flow. 
The present data can hardly discriminate between the two scenarios. 
Figure 6 reports also the values for  Mrk 231 (Feruglio et al. 2015) and IRAS F11119 (Tombesi et al. 2015). 
These are the only other two sources for which both the nuclear ultra-fast outflow  (UFOs) and the molecular wind have been measured.  

As for the loading factor, $\eta=\dot M_{OF}/SFR$, we note that the SFR estimates in APM 08279 are also very uncertain, because the FIR SED is dominated by the AGN emission (90\% due to the AGN). Weiss et al. (2007) derives a SFR of 25 $M_{\odot}~yr^{-1}$, while Riechers et al. (2009) suggests a much larger value of about 200 $M_{\odot}~yr^{-1}$. In all cases $\eta >> 1$, suggesting that the consumption rate of the molecular gas is driven by the outflow rather than by star formation.

\subsection{Emission line at 94.83 GHz}\label{uif}

\begin{table*}
\caption{Molecular species expected at frequencies close to 94.83 GHz}
\begin{center}
\begin{tabular}{lcccccc}
\hline
\multicolumn{1}{l} {Species}&
\multicolumn{1}{c} {Chem. Name}&
\multicolumn{1}{c} {Rest Frame Freq.}&
\multicolumn{1}{c} {Obs. Freq.}&
\multicolumn{1}{c} {Resolved QNs}&
\multicolumn{1}{c} {$E_L$} &
\multicolumn{1}{c} {Database} \\
&   & [GHz] & [GHz] &  & [K] &  \\
\hline
 	SO$_2$ v = 0             &	Sulphur dioxide 	     &  465.75116  & 94.81905  &  26(0,26)-25( 1,25)    & 	284.6362  &	CDMS\\
	HC$_3$N v6=1,v7=1 &	Cyanoacetylene             & 465.76703   & 94.82228  &  J=51-50, l= 0$^+$    &	 0.0000     &	CDMS\\	
 	N$_2$H$^+$ v = 0     &	Diazenylium 	    	     & 465.82478   & 94.83404  &   J= 5 - 4 	           &	44.7141    &	SLAIM\\
 	SO$_2$ v = 0             & 	Sulphur dioxide 	     & 465.88182   & 94.84565  &  25(10,16)-26( 9,17)  & 	520.9965  & 	CDMS\\
\hline
\end{tabular}
\end{center}
\label{tabid}
\end{table*}

We discuss in the following the identification of the 94.83 GHz feature. 
Table 5 lists the main molecular emission lines expected around that frequency. 
Identification with $\rm HC_3N $ rotational transition of the vibrationally excited (v7=1, v6=1) state seems excluded based on the non detection of the corresponding rotational line J=51-50 (rest frame frequency 463.7 GHz), which should be approximately a few times brighter (Martin et al. 2011, 2016 on Arp220, Sakamoto et al. 2010).

The recommended identification based on CDMS is N$_2$H$^+$(5-4) (diazenylium). 
The first extragalactic (and high-z) detection of N$_2$H$^+$ 
was provided by Wiklind \& Combes (1996) in absorption 
toward the lens galaxy of the background radio source PKS 1830-211.
They could identify $\rm N_2H^+$ based on the detection of two lines of the same species, the J=2-1 and the J=3-2.

In APM 08279+5255 we do not have two or more different rotational lines, and the line width is too large to use the N$_2$H$^+$ hyperfine structure to identify it. 
An alternative, more approximate possibility to identify the feature is given by the ratio HCO$^+$/N$_2$H$^+$. 
Since both HCO$^+$ and N$_2$H$^+$ are ions, we might expect that they are present
in similar molecular clouds irradiated by cosmic ray particles (e.g. Ceccarelli et al. 2014).
This ratio might be better than the ratio to CO, because the CO might
be coming from a completely different volume of gas than the $\rm N_2H^+$.
We might also expect that the HCO$^+$ to N$_2$H$^+$ ratio stays constant, as we go up the rotational ladder.
The candidate line in APM 08279+5255 is the J= 5-4, and this is the same transition that was observed in HCO$^+$ by
Garcia-Burillo et al. (2006).
They measured a HCO$^+$(5-4) line intensity of $0.87\pm 0.13$ Jy km s$^{-1}$, with a peak intensity of about 2.8 mJy/beam, 
with a large uncertainty on the line width (490 km s$^{-1}$) due to a noisy spectrum.
So if we tried N$_2$H$^+$ as a candidate for the feature in APM 08279+5255, then its (5-4) flux would be about $0.84$ mJy (Table 1), implying a HCO$^+$(5-4) / N$_2$H$^+$(5-4) ratio of 3.3 (at the line peak). 
We can now compare the HCO$^+$ to N$_2$H$^+$ ratios measured in nearby galaxies by Aladro et al. (2015). 
For NGC1068, the HCO$^+$/N$_2$H$^+$ ratio is 7.2 (integrated), or 7.4 (peak), in the (1-0) lines. 
For Arp220  the HCO$^+$/N$_2$H$^+$ ratio is 2.0 (integrated) or 1.7 (peak), in the (1-0) lines.
Similarly, in M51, the ratio is about 4-to-1, whereas  M82 seems to behave very differently, the ratio being  20-to-1 (integrated), or 12-to-1 (peak).
In NGC 253, the ratio is about 5-6. In M83, about 8.
If we ignore M82, the average ratio seems to be about 5, at least in the 1-0 lines.
So if the ratio stays the same, as we go up the ladder to (5-4)
in both ions, then that is what we would expect in APM 08279+5255.  

An alternative identification is sulphur dioxide, SO$_2$. 
Sulphur-bearing species are common in hot gas, and thought to be released from dust grains in shocked regions, which we likely have in APM 08279+5255, given the presence of fast outflows. 
In particular SO$_2$ abundance is found to be enhanced both within shocks and in extended post-shock regions  of interstellar clouds (Pineau des Forets et al. 1993). 
White et al. (2003) provided a spectral scan of the Orion-KL hot cloud core in the 455-507 GHz frequency range. 
There, no line is found at the 
frequency of N$_2$H$^+$, while there is a strong cluster of SO$_2$ lines at that right frequency, of which the brightest are SO$_2$ 26(0,26)-25(1,25) (45.3 K), and SO$_2$ 29(2,28)-29(1,29) (13.0 K). 
There is also NH$_2$CHO (22-21) (16.1 K). 
These features would be blended in our APM 08279+5255 spectrum. 
In Orion the ratio CO(4-3)/SO$_2$-blend is about 11 (White et al. 2003). 
In APM 08279+5255 the ratio CO(4-3)/SO$_2$ would be about 14 (using integrated intensities as in Table 1).
However, identification of this feature with SO$_2$ seems to be excluded, due to the absence of another transition at rest frame 455.77 GHz, which is found to be twice as bright in the Orion KL spectrum (White et al. 2003). 
In addition, if we assume that the abundance ratios are the same as in the Orion (Schilke et al. 2011) and in other hot cores (Beuther et al. 2009), that would imply the presence of other emission bands. 

Overall, N$\rm _2H^+$(5-4) from the QSO host galaxy seems the most viable identification. 
Unambiguous identification of the emission line, however,  
requires additional observations of other rotational transitions of the same species. 

Both $\rm N_2H^+$  and SO$_2$ species are commonly detected in the MW star forming regions (Orion KL), and in nearby galaxies (Martin et al. 2011, Aladro et al. 2015), but they were never detected before at such high redshift. 
We note that $\rm N_2H^+$ is a tracer of the ISM ionisation by cosmic rays (CR) in UV, and even X-ray opaque regions. 
SO$_2$ instead, as all sulphur bearing molecules, traces warm regions within shock or in the post-shock regions, as it is released from grain surfaces in the gas phase by shocks. 
Both such physical conditions, dense UV and X-ray opaque regions, and shocked regions, are likely present in APM 08279+5255.

An alternative possibility is that the emission feature is a CO emission line due to an intervening galaxy. 
This could also be the lens galaxy, which remains undetected in all observations available so far (Oya et al. 2013 and references therein).
The position of the emission (i.e. within 0.2\arcsec~ from the QSO image position) is compatible with the Egami et al. (2000) model, that requires the lensing potential centre to be almost on the line of sight to the QSO. 
They suggested that the lens is a massive galaxy at $\rm z\sim3$ with a dynamical mass of $2\times 10^{11}~\rm M_\odot$, core radius 0.2\arcsec, and located  in the redshift range z$=$0.5-3.5.
The core radius of 0.2\arcsec~ corresponds to physical scales of 1.2-1.5 Kpc at z=0.5 and 3.5, respectively. 
Similar values are proposed by Krips et al. (2007). 
Riechers et al. (2009) have pointed out that a core radius of 1.5 kpc seems too large for an elliptical at $z\sim3$, and has proposed a model with a spiral disc lens galaxy, which in turn implies a small magnification factor.  

Assuming that the line is either CO(4-3), CO(3-2) or CO(2-1), these would correspond to a redshift of the lens galaxy of z=3.862, 2.646 and 1.431, respectively, which are broadly consistent with all proposed lensing models. 
A redshift 0.21 galaxy derived in the case of CO(1-0) can be safely excluded, because it would have been detected in the  Subaru observations of Oya et al. (2013), if not also in previous shallower observations. 
However, the Oya et al. (2013) observations are limited to 17 mag in the L'-band and 15.3 mag in M'-band (5$\sigma$, point source), therefore would not detect a galaxy of mass $2\times 10^{11}~\rm M_\odot$ at redshift $\sim$1-3. 
We computed the H$_2$ masses for the three possible redshifts of the lens galaxy, by assuming the MW conversion factor. 
We find M(H$_2$) of the order $10^{10}~\rm M_\odot$. Given the dynamical mass of $2\times 10^{11}~\rm M_\odot$, that would imply a gas fraction of a few 10\% (considering all the uncertainties in conversion factors, excitation correction, etc). This value is compatible with the gas fraction measured for SF galaxies at z$\sim1-3$. To have a clearer picture requires confirmation by detection of other CO transitions from the candidate lens - which seems feasible with NOEMA.


\section{Conclusions}
We have performed a high sensitivity observation of the UFO/BAL quasar APM 08279+5255 at z=3.912 with NOEMA at 3.2 mm, aimed at detecting fast moving molecular gas. 
APM 08279 exhibits AGN-driven outflows acting at different scales and involving different gas phases (Chartas et al 2009, Saez \& Chartas 2011, Saturni et al. 2016, Hagino et al. 2017). 

Our findings are summarised below. 

(i) We detected a blueshifted CO(4-3) emission line component, 
with maximum velocity (v95\%) of $-1340$ km s$^{-1}$, with respect to the systemic velocity (peak of CO(4-3)), 
and spatially offset by $\sim1.5\arcsec$ from the narrow CO(4-3) emission peak. 
This blue wing has  a luminosity of 
$L' = 9.9\times 10^9 ~\mu^{-1}$ K km s$^{-1}$ pc$^{-2}$ ($\mu$ is the magnification factor). 
We interpret this as the first detection of molecular gas with high velocity at redshift $>3$, tracing an outflow of molecular material arising from the front side of the QSO. 
Yet, we caution that this emission could also have a alternative nature, the main possibilities being (a) inflowing molecular gas from the back side of the QSO, and (b) a separate merging or companion galaxy. Higher resolution observations are required to assess the nature of this emission.

(ii) We derived a 
mass flow rate of the fast CO gas of $3-7.4\times 10^3$ M$_\odot$/yr, and momentum boost of the molecular wind $\dot P_{OF} / \dot P_{AGN}  = 2-6$. 
The momentum boost of the molecular wind is therefore within the range of values compatibles with a momentum driven flow. 
For the largest $\dot P_{OF}$ the scaling is also consistent with a energy conserving flow with an efficiency of $\sim$10-20\%. 
The present data can hardly discriminate between the energy- and momentum conserving scenarios. 
The mass loading factor of the molecular outflow $\eta=\dot M_{OF}/SFR$ is in all cases $>>1$, suggesting that the consumption rate of the molecular gas is driven by the outflow rather than by star formation.

(iii) We detected a molecular emission line at a frequency of 94.83 GHz, corresponding to a rest frame frequency of 465.8 GHz, which we tentatively identified with the cation molecule $\rm N_2H^+$(5-4), which would be the first detection of this species at high redshift. Additional observations are required to confirm it, by detecting at least one other transition.

New useful insights on the nature of the emission blueward of the CO(4-3) line may be obtained by observing with NOEMA at other frequencies and/or with higher angular resolution. These have the potential to confirm the detection, constrain the size of the emitting regions, and to provide clues on the excitation of the gas. The latter could be instrumental to ascribe the emission to a galaxy or to outflowing/inflowing gas.
A major improvement in our understanding of the UFO-galactic wind connection is expected from the ALMA observations of PDS456, a nearby (z=0.18) QSO, which is the best studied nuclear wind, and free from all the shortcomings of gravitationally lensed sources.

\begin{acknowledgements}
The authors thank IRAM Astronomy Group Staff for making these observations possible. 
We thank S. Garcia-Burillo, Chris Done, George Chartas, Fabrizio Nicastro and Francesco Tombesi for providing interesting input. 
We thank Paolo Fiore for suggestions to improve the graphical presentation of Figure 2.
CF acknowledges support from  the European Union Horizon 2020 research and innovation programme under the Marie Sklodowska-Curie grant agreement No 664931, and from MIUR 2010-2011 prot. 2010LY5N2T. 
FF acknowledges financial support from INAF under the contract PRIN-INAF-2011. 
RM acknowledges support by the Science and Technology Facilities Council (STFC) and ERC Advanced Grant 695671 "QUENCH". 
EP acknowledges financial support from INAF under the contract PRIN-INAF-2012. 
LZ acknowledges financial support under ASI/INAF contract I/037/12/0
\end{acknowledgements}

 \end{document}